# Group A Rotavirus NSP4 is Under Negative Selective Pressure.

Authors:

Jackson Cordeiro Lima[1], Paulo Bandiera-Paiva[1] *

[1] Health Informatics Department, Universidade Federal de São Paulo

Rua Botucatu, 862

CEP 04023-062

São Paulo, SP

Brazil.

**Running title: Rotavirus NSP4 under negative pressure**

* **Corresponding author**:

Telephone: +55 11 5576-4347

E-mail: **paiva@unifesp.br**

**Abstract:**

Rotavirus (RV) is the major etiologic agent of severe infantile gastroenteritis; its genome has 11 segments of double stranded RNA, encoding 12 proteins. The non-structural protein 4 (NSP4) encoded by segment 10 is multifunctional. The aim of this study is to analyze the selective pressure driving the NSP4 of RV, through the ratio of non-synonymous substitutions per synonymous substitutions (dN/dS). Our results show that NSP4 is under negative evolutionary pressure (84.57% of the amino acid sequence) and no site was found under positive selection. This may support other evolutionary studies of different RV proteins or viral agents.

# Introduction:

Group A Rotavirus (RV) are the major causes of gastroenteritis of viral origin in children with less than five years of age, 611,000 deaths a year caused by severe diarrhea due to RV are estimated. Most of these occur in developing countries (Parashar et al., 2006). RV belongs to the family *Reoviridae*, a non-enveloped virus with three concentric layers of protein that enclose a genome composed of 11 double stranded RNA (dsRNA) segments, which encode 6 structural proteins (VP1-VP4, VP6 and VP7) and 6 nonstructural proteins (NSP1-NSP6) (Estes, 2001).

The nonstructural protein 4 (NSP4) encoded by segment 10 has 175 amino acids and 3 hydrophobic domains: H1 in the lumen of the endoplasmic reticulum (ER), H2 which is a ER transmembrane and H3 in the C-terminal localized on the cytoplasm. The C-terminal portion of the protein is a receptor to the immature viral capsid which has an oligomerization region at amino acids 95-133 and a calcium-binding site (Huang et al., 2004). Phylogenetic analyses reported that NSP4 of group A RV are classified in 5 genotypes A-E based on prototype references, KUN, WA, AU-1, EW and Ch-1, respectively (Ciarlet et al., 2000; Iturriza-Gómara et al., 2003; Lin and Tian, 2003).

NSP4 is a multifunctional protein; it became well known in 1996 due to its role in the pathogenesis of RV and was classified as the first viral enterotoxin (Ball et al., 1996). Other viral toxins have been reported (Ball et al., 2005b). The critical role in the morphogenesis process of RV was demonstrated. Being a transmenbrane protein, it is responsible for the passage of the immature virus from the cytoplasm to the ER, where the virus acquires its third proteic layer (Ball et al., 2005a).

Studies reported that NSP4 is well conserved among the different genotypes and that there are no motifs in NSP4 which differentiate the virulent from the non-virulent samples (Lee et al., 2000; Oka et al., 2001). The region among the amino acids 131-140, located in the connection with VP4 and VP6, presents high diversity (Lin and Tian, 2003; Jagannath et al., 2006). Due to its characteristics, relevance and lack of evolutionary studies involving NSP4, our work aims to understand why NSP4 is well conserved and how the evolutionary forces are driving it.

## Material and Methods:

### Sequence Analysis:

The dataset used in this study comprises 90 sequences of NSP4 from group A rotavirus, including the 5 genotypes, it was obtained from GenBank (NCBI v157) searches followed by a refinement which considered two criteria: only complete open reading frames and sequences which have their collection date referred in literature.

The dataset was aligned using the MAFFTv5 (Katoh et al., 2005), using the methodology *G-insi*, followed by a manual revision of the alignment, in order to prevent gaps in the middle of the codons.

### Phylogenetic analysis and Positive Selection:

Phylogenetic analysis was accomplished in three steps using the HYPHY program v0.9999 (Pond et al., 2005). A tree was built by Neighbor-Joining with Maximum Likelihood (Felsenstein, 1981), using the General Time Reversible (GTR) model (Tavaré, 1986). The model of nucleotide substitution which best fits the data was selected by Modeltest (Posada and Crandall, 1998), using the Akaike Information Criterion (AIC) (Akaike, 1974). A new tree was inferred with the selected model and parameters obtained from Modeltest.

The detection of the selective pressure acting upon NSP4 was performed by HYPHY, using two methodologies: Single Likelihood Ancestor Counting (SLAC), and Fixed Effects Likelihood (FEL) (Kosakovsky-Pond and Frost, 2005), with a p-value of 0.05 for the detection of positive sites.

## Results:

The best nucleotide substitution model found was the General Time Reversible (GTR) using Gamma (4 classes), with the rates of AC = 0.14457, AG=1.0, AT=0.22819, CG=0.04479, CT=1.27727 and GT=0.10168, the nucleotide frequencies of A=0.3831, C=0.1568, G=0.1885, T=0.2717.

The pattern of evolutionary pressure on NSP4 is negative, or purifying, from the 175 sites of the protein, 137 (78.28%) are under negative evolutionary pressure using the SLAC methodology and 38 (21.72%) are under neutral pressure, no site was found to be under positive selection (Figure 1A). Using the FEL methodology, 148 sites were found under negative evolutionary pressure, while 27 sites developed in neutral pressure, no site under positive selection was found (Figure 1B).

The global value of non-synonymous per synonymous substitutions (dN/dS) found using the FEL methodology was of 0.102723 and a Log Likelihood value of -14593.9974398711, showing a low rate of non-synonymous substitutions, corroborating that the evolutionary selection is negative.

Most of the non-synonymous substitutions don't alter the hydrophobic/hydrophilic character of the amino acids. When there is a modification of the amino acid character, it usually changes to one with a similar character, e.g., from polar neutral to apolar.

The enterotoxic domain of the protein, amino acids 114-135, was found to be very conserved, only two amino acids (119 and 135) were under neutral evolution.

The amino acids 131-140 are described in the literature as presenting a high degree of variation. This segment includes the interspecies-variable domain, amino acids 135-141, which presents amino acids 135-139 as developing under neutral selection.

The immature capsid coupling site, amino acids 161-175, also presents a conservation tendency, and possibly, its connection to VP6 takes place in a negative evolutionary pressure region, which corresponds to 27% of the protein (not shown).

The hydrophobic domains H1, H2 and H3, showed that H3, which is located in the host organisms cytoplasm, tends to be more conserved, only 2 amino acids (70 and 72) are under neutral selective pressure.

The VP4 binding site, amino acids 112-146, showed a conservation tendency, presenting 26% of the amino acids under neutral pressure.

## Discussion:

The analysis of selective pressure shows how the protein or organism evolves along time, in this manner, using phylogenetic tests and molecular clock it is possible to confirm its presumed evolutionary history. This is the first evolutionary footprint study involving the NSP4 of Group A rotavirus.

The data analysed in this study suggests that NSP4 is under negative evolutionary pressure, maintaining itself conserved since mutations in the protein can lead to non-infecting or structurally incomplete RV.

The negative pressure suffered by NSP4 possibly is related to its function in the viral transcription, replication and "genome packaging" (Xu et al., 2000; Silvestri et al., 2005). Studies using small interference RNA (siRNA) of rotavirus (Lopez et al, 2005) show that when using siRNA of NSP4 there was an inhibition of the formation of ripe particles of the virus. This may explain the negative pressure found in NSP4, since this protein has an important role in the morphogenesis process.

In spite of RNA viruses be known for having high substitution rates (Holmes, 2003; Chen and Holmes, 2006; Belshaw et al., 2007; Normile, 2007), data for NSP4 of RV suggests the protein cannot suffer mutations, being more conserved than the polimerase of RV, which possesses 37.86% of the amino acids under positive pressure, 11.39 % and 50.75% under negative and neutral evolutionary pressure (not shown).

As other studies suggest (Lin and Tian, 2003; Jagannath et al., 2006) the interspecies-variable region at amino acids 131-140 presents a high degree of variability. It may be possible to find traces of positive selection in that region since rotavirus infect human, porcine, equine, felines, canines, murines and avian. Our results show that 6 amino acids suffer negative evolutionary pressure and 4 are under neutral selection.

This negative pressure may explain why prior NSP4 studies failed in their attempt to identify differences between symptomatic and asymptomatic children (Lin and Tian, 2003; Mascarenhas et al., 2007). The protein seems to have a narrow scope of possible mutations which keep its functionality.

Balasuriya et al. (2008) analyzed the evolutionary pressure of NS3 of bluetongue virus and found that NS3 presents 53% of the protein under negative pressure. This protein has a similar function to the NSP4; both are transmembrane and participate in the assembly of the viral particle.

These results may support other evolutionary studies with nonstructural proteins of RV or other viral agents.

## Acknowledgments:

JCL had a master fellowship granted by Coordenação de Aperfeiçoamento de Pessoal de Nível Superior (CAPES) (Brazil).

**Figure and Legend**

Figure 1 – Schematic linear representation of NSP4 with the location of functional domains and the selective pressure of each site. White represents sites with neutral evolutionary pressure and gray a negative evolutionary pressure. (A) Results of SLAC methodology and (B) results of FEL methodology. The three hydrophobic domains: H1 (amino acids 7-21), H2 (amino acids 28-47) and H3 (amino acids 67-85). ER transmembrane domain (amino acids 24-44). Amphipathic alpha helix and Coiled Coil Domain (amino acids 95-137). Enterotoxic Pepdtide (amino acids 114-135), Interspecies variable domain (amino acids 135-142), VP4 binding (amino acids 112-148) and VP6 binding (amino acids 161-175) (Bowman et al., 2000; Ball et al., 1996; Lin and Tian, 2003).

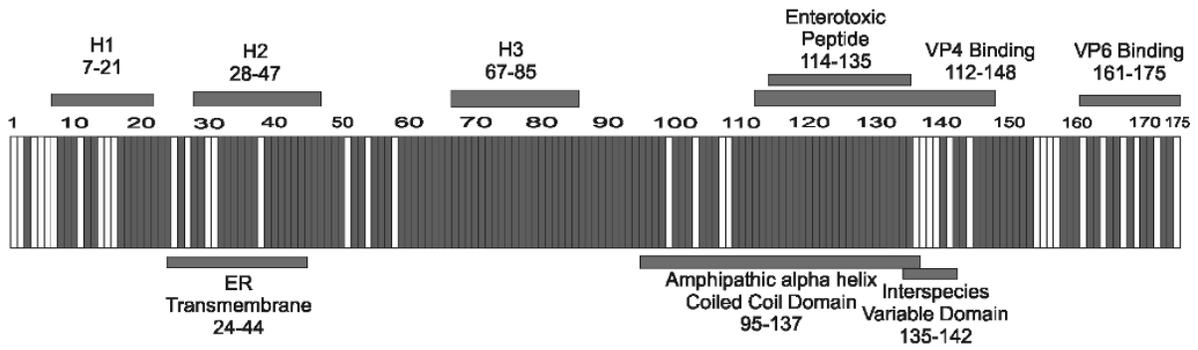
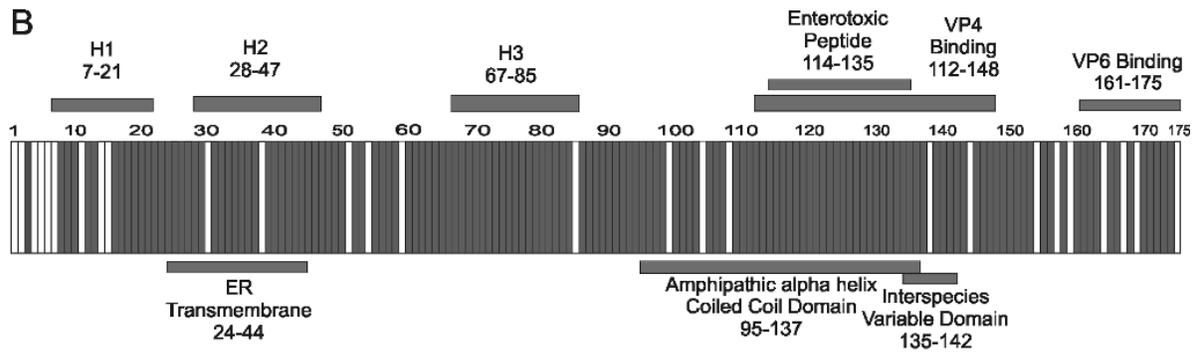